\documentclass{jfm}

\usepackage{graphicx}
\usepackage{natbib}
\usepackage{epstopdf}
\usepackage{amsmath}
\usepackage{esdiff} 
\usepackage{subfig}
\usepackage{float}

\ifCUPmtlplainloaded \else
  \checkfont{eurm10}
  \iffontfound
    \IfFileExists{upmath.sty}
      {\typeout{^^JFound AMS Euler Roman fonts on the system,
                   using the 'upmath' package.^^J}%
       \usepackage{upmath}}
      {\typeout{^^JFound AMS Euler Roman fonts on the system, but you
                   dont seem to have the}%
       \typeout{'upmath' package installed. JFM.cls can take advantage
                 of these fonts,^^Jif you use 'upmath' package.^^J}%
      }
  \else
  \fi
\fi

\ifCUPmtlplainloaded \else
  \checkfont{msam10}
  \iffontfound
    \IfFileExists{amssymb.sty}
      {\typeout{^^JFound AMS Symbol fonts on the system, using the
                'amssymb' package.^^J}%
       \usepackage{amssymb}%
       \let\le=\leqslant  \let\leq=\leqslant
         
      }{}
  \fi
\fi

\ifCUPmtlplainloaded \else
  \IfFileExists{amsbsy.sty}
    {\typeout{^^JFound the 'amsbsy' package on the system, using it.^^J}%
     \usepackage{amsbsy}}
    {\providecommand\boldsymbol[1]{\mbox{\boldmath $##1$}}}
\fi

\providecommand\bnabla{\boldsymbol{\nabla}}

\newsavebox{\astrutbox}
\sbox{\astrutbox}{\rule[-5pt]{0pt}{20pt}}

\title{Transient Dynamics of Elastic Hele-Shaw Cell Due to External Forces with Application to Impulse Mitigation}
\shorttitle{Transient Dynamics of Elastic Hele-Shaw Cells}
\author[A. Tulchinsky and A.D. Gat]{A. Tulchinsky and A.D. Gat}
\affiliation{Faculty of Mechanical Engineering, Technion - Israel Institute of Technology, Haifa 3200003, Israel}

\pubyear{2015} \volume{999} \pagerange{999--9999}
\date{2015}
\begin{document}
\maketitle

\abstract{We study the transient dynamics of a viscous liquid contained in a narrow gap between a rigid surface and a parallel elastic plate. The elastic plate is deformed due to an externally applied time-varying pressure-field. We model the flow-field via the lubrication approximation, and the plate deformation by the Kirchhoff-Love plate theory. We obtain a self-similarity solution for the case of an external point force acting on the elastic plate. The pressure and deformation field during and after the application of the external force are derived and presented by closed form expressions. We examine a uniform external pressure acting on the elastic plate over a finite region and during a finite time period, similar to the viscous-elastic interaction time-scale. The interaction between elasticity and viscosity is shown to reduce by order of magnitude the pressure within the Hele-Shaw cell compared with the externally applied pressure, thus suggesting such configurations may be used for impact mitigation.}

\section{Introduction} 
We study the transient dynamics of viscous liquid contained in the narrow gap between a flat rigid surface and a parallel elastic plate, subjected to external time-varying pressure field. In such configurations, external energy applied to the system does not immediately dissipates owing to the creation of elastic potential energy, yielding viscous-elastic transient dynamics. 




Recent works on external force induced fluid-structure dynamics include \cite{vella2015indentation}, who studied the formation and evolution of wrinkles due to indentation of a thin elastic plate floating over a liquid surface. \cite{duchemin2014impact} examined the dynamics due to impact of a rigid body on a thin elastic sheet laying over a liquid. \cite{FLM:9795348} studied externally forced deformation of a saturated poroelastic medium laying on rough wet surface in the context of adhesion.

Works involving viscous flow in the gap between an elastic plate and a rigid surface include \cite{chauhan2002settling}, who examined a configuration of an elastic shell positioned over a viscous film laying on a rigid surface in the context of contact lens dynamics during blinking. Corneal air puff tests, which examine the intraocular pressure via an air jet applied on the eye \citep{han2014air}, also include a similar configuration of an external pressure applied on an elastic shell laying over a liquid film. \cite{hosoi2004peeling} studied the peeling and busting dynamics of a viscous liquid contained between a rigid surface and an elastic sheet. Taylor-Saffman instability in elastic Hele-Shaw cell configurations was examined by \cite{pihler2012suppression} and \cite{al2013two}, who showed that elasticity delays the onset of the instability. \cite{pihler2014interaction} related the patterns of viscous fingering to patterns of wrinkling of the elastic Hele-Shaw cell.  \cite{trinh2014pinned} and \cite{2014arXiv1410.8558T} studied a rigid and elastic plates, either pinned or free floating, moving over a viscous film laying on a flat rigid surface. The displacement dynamics of a liquid entrained between an elastic sheet and a rigid surface, due to injection of another fluid, was studied experimentally, numerically and analytically by \cite{peng2015displacement} and \cite{pihler2015displacement}.

The aim of this work is to relate the externally applied pressure field to the elastic deformation and pressure distribution created within the Hele-Shaw cell, during and after the application of an external pressure field. Specifically, we examine pressure impulse mitigation properties of such configurations. This work is arranged as follows: In \S2 we present the problem formulation. In \S3 we obtain relevant Green functions and self-similar solutions for the governing equations. In \S4 we examine the system dynamics during and after application of external forces and the impulse mitigation properties of such configurations. 

\section{Problem Formulation}
\begin{figure}
\centering 
\includegraphics[width=0.9\textwidth]{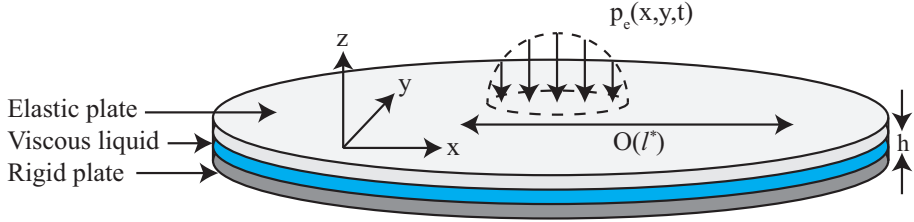}
\caption{Illustration of the configuration and coordinate system. The plates are parallel to the $x-y$ plane. $p_e (x,y,t)$ is the external pressure field and $h$ is the gap between the rigid surface and the elastic plate.}
\label{figure1}
\end{figure}
We study transient creeping flow in the narrow gap between a rigid surface and a parallel elastic plate due to time-varying external pressure acting on the elastic plate. The configuration and coordinate system are defined in Fig. \ref{figure1}. Hereafter, asterisk superscripts denote characteristic values and Capital letters denoted normalize variables. The subscript $\perp$ denotes the vector component perpendicular to the $x-y$ plane and no subscript denotes the two-dimensional vector components parallel to the $x-y$ plane.

The gap between the surfaces is $h$, the deformation of the plate is $d$, the plate bending resistance is $s$, the plate width is $b$, the plate density is $\rho_s$, the liquid pressure is $p$, the liquid viscosity is $\mu$, the liquid density is $\rho_l$, the fluid velocity is $(\boldsymbol{u},u_\perp)$, and the external pressure field is $p_e$.


The characteristic length scale in the $x-y$ plane is $l^*$, the characteristic pressure is $p^*$, the initial liquid film height is $h_0$, the characteristic deformation is $d^*$, the characteristic speed in the $x-y$ plane is $u^*$ and the characteristic speed in the $z$ direction is $u_\perp^*$. The characteristic stress resultant acting perpendicular to $z$ direction is $n^*$.


We  define the small parameters 
\begin{equation}
\varepsilon_1=\frac{h_0}{l^*}\ll1, \quad
\varepsilon_2=\frac{d^*}{h_0}\ll1,\quad \frac{\varepsilon_1}{\varepsilon_2}\frac{\rho_l u^* h_0}{\mu}\ll1,\quad\frac{\rho_s b l^{*4}}{t^{*2} s}\ll1,\quad \frac{n^* l^{*2}}{s}\ll1
\label{assumption_eq}
\end{equation}
corresponding to assumptions of shallow geometry, small ratio of transverse plate deformations to viscous film height, negligible fluid inertia, negligible solid inertia and negligible membrane effects, respectively. Under the assumptions given in (\ref{assumption_eq}), the Hele-Shaw cell's upper elastic plate dynamics are governed by the Kirchhoff-Love equation
\begin{equation}
-s\nabla^4 d+p-p_e=0,
\label{love}
\end{equation}
and a boundary condition requiring that sufficiently far from the location of the external force the deformation vanishes
\begin{equation}
d(|\boldsymbol{x}|\rightarrow\infty) \rightarrow 0.
\end{equation}
The Newtonian, incompressible fluid located within the elastic cell is governed by the continuity 
\begin{equation}
\nabla \cdot \boldsymbol{u}=0,
\end{equation}
and momentum equations
\begin{equation}
\rho_l\left(\frac{\partial \boldsymbol{u}}{\partial t}+\boldsymbol{u}\cdot\nabla\boldsymbol{u}\right)=-\nabla p +\mu \nabla^2\boldsymbol{u},
\label{momentum_eq}
\end{equation}
with the boundary condition that pressure is uniform sufficiently far from the location of the external pressure
\begin{equation}
\bnabla{} p(|\boldsymbol{x}|\rightarrow\infty) \rightarrow 0,
\end{equation}
as well as no-slip and no-penetration at the solid-liquid interfaces
\begin{equation}
(\boldsymbol{u},u_\perp)(z=-h)=(\boldsymbol{0},0),\quad u_\perp(z=d)=\frac{\partial d}{\partial t} +\boldsymbol u \cdot \bnabla d,\quad
\boldsymbol u(z=d)=-\frac{b}{2}\bnabla{}
\left(\frac{\partial d}{\partial t}\right),
\end{equation} 
where the term $(b/2)\bnabla (\partial d/\partial t)$ represents in-plane velocities due to angular speed. 

We define the normalized variables
\begin{equation}
P=\frac{p}{p^*},\quad (\boldsymbol{X},X_\perp)=\left(\frac{\boldsymbol{x}}{l^*},\frac{x_\perp}{h_0}\right),\quad (\boldsymbol{U},U_\perp)=\left(\frac{\boldsymbol{u}}{u^*},\frac{u_\perp}{u_\perp^*}\right),\quad D=\frac{d}{d^*},
\label{normalized_variables}
\end{equation}
corresponding to the normalized pressure $P$, coordinates $(\boldsymbol{X},X_{\perp})$, fluid velocity $(\boldsymbol{U},U_\perp)$, and solid deformation $D$. Substituting (\ref{normalized_variables}) into (\ref{momentum_eq}) yields, in leading order 
\begin{equation}
\begin{gathered}
\bnabla{} P=\frac{1}{12}\bnabla{^2\boldsymbol U}+O\left(\frac{\varepsilon_1}{\varepsilon_2}Re,\varepsilon_1^2\right), \quad
\bnabla{}_\perp P=O\left(\frac{\varepsilon_1^3}{\varepsilon_2}Re,\varepsilon_1^2,\right)
\end{gathered}
\label{normalized_momentum}
\end{equation}
and 
\begin{equation}
\frac{\partial U_\perp}{\partial Z}+\bnabla\cdot\boldsymbol{U}=0,
\label{normalized_mass}
\end{equation}
where order of magnitude yields $p^*=12\mu u^*l^*/h_0^2$ and $u^*=u_\perp^*/\varepsilon_1$. Substituting (\ref{normalized_momentum}) into (\ref{normalized_mass}), integrating with respect to $Z$, 
\begin{equation}\label{deft_to_press_mass}
\frac{\partial D}{\partial T}+\bnabla^2P=O\left(\varepsilon_1\varepsilon_2\frac{b}{l^*}\right),
\end{equation}
where order of magnitude analysis yields $u_\perp^*=d^*/t^*$. Substituting (\ref{love}) into (\ref{deft_to_press_mass}) yields the governing equation in term of deflection $D$
\begin{equation}
\frac{\partial D}{\partial T}-\bnabla^6 D=\bnabla^2 P_e
\label{D_gov}
\end{equation}
with boundary condition $D(\boldsymbol{X}\rightarrow\infty)\rightarrow0$. Alternatively, the governing equation can be presented with regard to the pressure
\begin{equation}
\frac{\partial P}{\partial T}-\bnabla^6 P=\frac{\partial P_e}{\partial T}
\label{P_gov}
\end{equation}
with boundary condition $P(\boldsymbol{X}\rightarrow\infty)\rightarrow0$.


\section{Green Functions and Self-Similarity}
The Green function for the $6^{th}$-order evolution equations (\ref{D_gov}) and (\ref{P_gov}) is given by \cite{satsanit2009operator} as
\begin{equation}
G=\frac{1}{4\pi^2}\int_{\mathbb{R}^2} e^{-(T-\bar{T}) |\boldsymbol\lambda|^6+i\boldsymbol\lambda\cdot(\boldsymbol X-\bar{\boldsymbol X})}d\boldsymbol{\lambda},\quad T>\bar{T}
\label{G_sol}
\end{equation}
where  $\bar{\boldsymbol X}$ and $\bar{T}$ are the location and time of the delta function, respectively. Eq. (\ref{G_sol}) represents the solution for the evolution of the pressure for external pressure, $\partial P_e/\partial T=\delta(T-\bar{T})\delta(\boldsymbol{X}-\bar{\boldsymbol{X}})$, and thus $P_e=\theta(T-\bar{T})\delta(\boldsymbol{X}-\bar{\boldsymbol{X}})+C_1(\boldsymbol{X})$. Similarly, (\ref{G_sol}) represents the solution for the evolution equation of the deformation (\ref{D_gov}) for $\partial^2P_e/\partial X^2=\delta(T-\bar{T})\delta(\boldsymbol{X}-\bar{\boldsymbol{X}})$ and thus $P_e=\delta(T-\bar{T})\theta(\boldsymbol{X}-\bar{\boldsymbol{X}})\boldsymbol{X}+X\cdot C_1(T)+C_2(T)$. We note that external pressure of the form $P_e=X\cdot C_1(T)+C_2(T)$ do not create deformation of the plate.

Eq. (\ref{G_sol}) may be interpreted as the inverse Fourier transform, where the argument of transformation is $e^{-(T-\bar{T})|\boldsymbol\lambda|^6}$. Furthermore, radial symmetry of the argument enables representation of (\ref{G_sol}) by the inverse Hankel transform
\begin{equation}
G=\frac{1}{2\pi}\int_{0}^{\infty}e^{-(T-\bar{T})\rho^6}J_0(\rho |\boldsymbol X-\bar{\boldsymbol X}|)\rho d\rho.
\end{equation}
Expressing the Bessel function in a series form, and integrating each element according to 
\begin{equation}
\int_0^\infty l^m e^{-\beta l^n}dl=\frac{\Gamma(\gamma)}{n\beta^\gamma},\,\,\,\gamma=\frac{m+1}{n},
\end{equation}
yields a self-similar expression
\begin{equation}\label{G_similar}
G=\frac{\Psi(\eta)}{(T-\bar{T})^{\frac{1}{3}}},\quad\eta=\frac{|\boldsymbol{X}-\bar{\boldsymbol X}|}{6 (T-\bar{T})^\frac{1}{6}}
\end{equation}
where
\begin{equation}
\Psi(\eta)=\frac{1}{12\pi }\sum\limits_{m=0}^{\infty}\frac{3^{2m}\Gamma(\frac{1+m}{3})(-1)^m}{\Gamma(m+1)^2}\eta^{2m}.
\end{equation}
We decompose the series into three separate series
\begin{equation}
\sum_{k=0}^{\infty}a_m \eta^k=\sum_{k=0}^{\infty}a_{3k} \eta^{3k}+\sum_{k=0}^{\infty}a_{3k+1} \eta^{3k+1}+\sum_{k=0}^{\infty}a_{km+2} \eta^{3k+2}
\end{equation}
thus yielding a closed-form expression in term of generalized hyper-geometric functions
\begin{multline}\label{sol_3}
\Psi(\eta)=\frac{1}{12 \pi}\bigg[\Gamma \left(\frac{1}{3}\right) \, _0F_4\left(;\frac{1}{3},\frac{2}{3},\frac{2}{3},1;-\eta^6\right)\\-9\eta^2 \Gamma \left(\frac{2}{3}\right) \, _0F_4\left(;\frac{2}{3},1,\frac{4}{3},\frac{4}{3};-\eta^6\right)+\frac{81}{4}\eta^4 \, _0F_4\left(;\frac{4}{3},\frac{4}{3},\frac{5}{3},\frac{5}{3};-\eta^6\right)\bigg].
\end{multline}

\begin{figure}
\centering
\includegraphics[width=\textwidth]{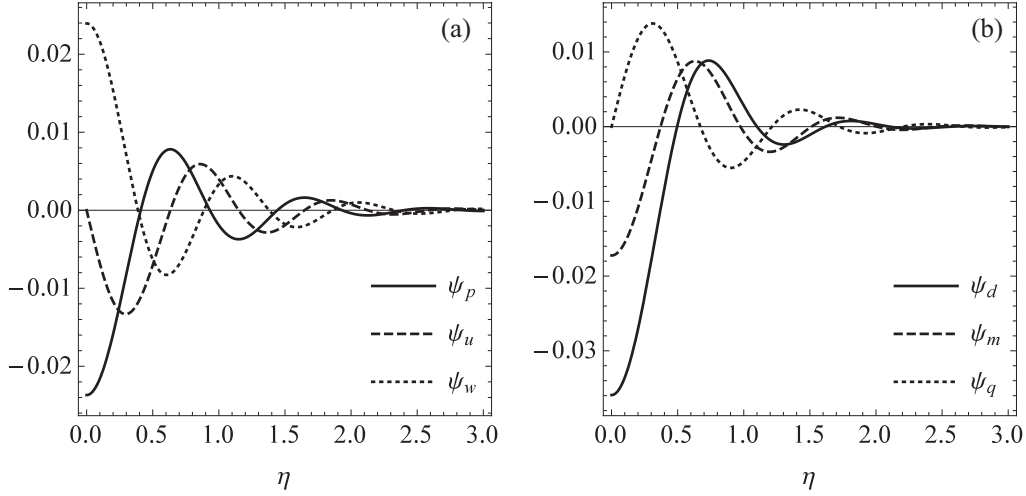}
\caption{The similarity shape function vs. $\eta$. Part a: pressure (solid), radial liquid velocity (dashed), transverse liquid velocity (dotted). Part (b) deformation (solid), bending moment (dashed) and transverse shear force (dotted).}
\label{figure2}
\end{figure}


While the function presented in (\ref{sol_3}) can be used by convolution to obtain a general solution, more insight may be obtained from a solution for the case of  $P_e=\delta(\boldsymbol{X}-\bar{\boldsymbol X})\delta(T-\bar{T})$. This may be achieved without convolution by applying the Laplacian operator in terms of $\boldsymbol{\bar{X}}$ (i.e. $\Delta_{\boldsymbol{\bar{X}}}$) on the equation defining the Green function. Linearity of the equation, and the relation $\Delta_{\boldsymbol{\bar{X}}}\delta(\boldsymbol{X}-\boldsymbol{\bar{X}})=\Delta\delta(\boldsymbol{X}-\boldsymbol{\bar{X}})$, yields
\begin{equation}
\left(\frac{\partial }{\partial T}-\nabla^6 \right)\Delta_{\boldsymbol{\bar{X}}}G=\Delta\delta(\boldsymbol{X}-\boldsymbol{\bar{X}})\delta(T-\bar{T}).
\end{equation}
Thus, the deformation-field due to a unit impulse is
\begin{equation}\label{G_d_to_G}
G_d=\Delta_{\boldsymbol{\bar{X}}}G=\frac{\Psi_d(\eta)}{(T-\bar{T})^\frac{2}{3}}
\end{equation}
where 
\begin{multline}
\Psi_d=-\frac{1}{48 \pi }\bigg[4 \Gamma \left(\frac{2}{3}\right) \,_0F_4\left(;\frac{1}{3},\frac{1}{3},\frac{2}{3},1;-\eta^6\right)\\+27 \eta^4 \Gamma\left(\frac{1}{3}\right) \,_0F_4\left(;1,\frac{4}{3},\frac{5}{3},\frac{5}{3};-\eta^6\right)-36 \eta^2 \, _0F_4\left(;\frac{2}{3},\frac{2}{3},\frac{4}{3},\frac{4}{3};-\eta^6\right)\bigg].
\end{multline}
A similar approach may be used to obtain the pressure-field due to a unit impulse, yielding
\begin{equation}\label{G_p_to_G}
G_p=-\frac{\partial G}{\partial \bar{T}}=\frac{\Psi_p(\eta)}{(T-\bar{T})^\frac{4}{3}}
\end{equation}
where
\begin{multline}
\Psi_p=-\frac{1}{36 \pi }\bigg[\Gamma \left(\frac{1}{3}\right) \, _1F_5\left(\frac{4}{3};\frac{1}{3},\frac{1}{3},\frac{2}{3},\frac{2}{3},1;-\eta^6\right)\\+\frac{243}{4} \eta^4 \, _1F_5\left(2;1,\frac{4}{3},\frac{4}{3},\frac{5}{3},\frac{5}{3};-\eta^6\right)-18 \eta^2 \Gamma \left(\frac{2}{3}\right) \, _1F_5\left(\frac{5}{3};\frac{2}{3},\frac{2}{3},1,\frac{4}{3},\frac{4}{3};-\eta^6\right)\bigg].
\end{multline}
 $G_p$, $G_d$ thus give more direct insight on the response to external forces and may be used to convolve $P_e$ directly for general solutions, similarly to a regular Green function. We note the characteristic pressure is $p_e^*=j_e^*/l^{*2}t^*$, where $j_e$ is the magnitude of the impulse.
 

 Without loss of generality, we assign $\bar{T}=0$. From $\eta$ , the radial speed of the signal propagation is obtained $\eta_{ref}T^{-5/6}$, where $\eta_{ref}$ is a reference state. Utilizing $G_p=T^{-4/3}\Psi_p(\eta)$ and $G_d=T^{-2/3}\Psi_d(\eta)$, the response to a localized external impulse can be calculated for the radial fluid speed $G_u=T^{-3/2}\Psi_u(\eta)$, transverse fluid speed $G_w=T^{-5/3}\Psi_w(\eta)$, bending moment  $G_m=T^{-1}\Psi_m(\eta)$ and shear force $G_q=T^{-7/6}\Psi_q(\eta)$. $\Psi_p$, $\Psi_u$, $\Psi_w$ are presented in Fig. \ref{figure2}a (solid, dashed and dotted lines, respectively) and $\Psi_d$, $\Psi_m$, $\Psi_q$ are presented in Fig. \ref{figure2}b (solid, dashed and dotted lines, respectively). All $\Psi_i$, ($i=p,u,w,d,m,q$) are similar decaying oscillating functions of $\eta$. However, a difference in the decay rate in time exists due to the different powers of $T$ multiplying $\Psi_i$. The slowest time decay is of the deformation, scaling as $T^{-2/3}$.

\section{Impact mitigation and response dynamics to spatially and temporally distributed external forces}
\begin{figure}
\centering
\includegraphics[width=\textwidth]{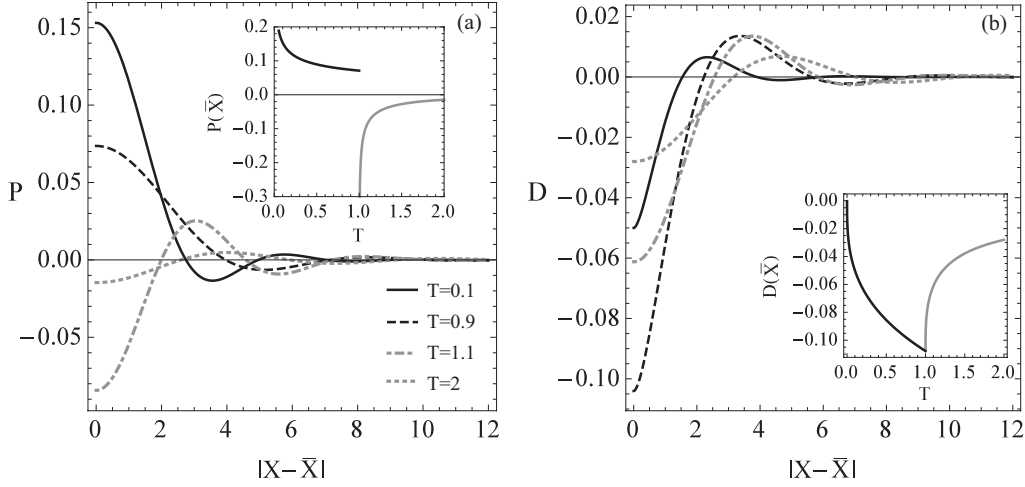}
\caption{Dynamics during and after application of a temporally uniform external force applied at $\boldsymbol{X}=\boldsymbol{\bar{X}}$. Parts (a), (b) show the pressure and deformation during $(T=0.1, T=0.9)$ and after $(T=1.1, T=2)$ application of the external pressure, respectively, where $T_e=1$. Inserts of (a), (b) present the time evolution of the pressure and deformation at the impact locus, receptively.}
\label{figure3}
\end{figure}

The functions (\ref{G_similar}), (\ref{G_d_to_G}) and (\ref{G_p_to_G}) can now be used to examine the fluid pressure-field and plate deformation-field created during and after application of spatially and temporally distributed external forces. For a temporally uniform external force applied at $\boldsymbol{X}=\boldsymbol{\bar{X}}$ over a finite time interval, defined as 
\begin{equation}\label{sudden_pe}
P_e=\frac{1}{T_e}\delta(\boldsymbol X-\bar{\boldsymbol X},T)[\theta(T)-\theta(T-T_e)],
\end{equation}
where $\theta$ is the Heaviside function and $T_e$ is the instant of release, the pressure-field is immediately obtained from (\ref{G_similar}) as
\begin{equation} \label{press_dueto_force}
  P=\frac{1}{T_e}\begin{cases}
    G(\eta,T), & T<Te\\
    G(T)-G(T-T_e), & T>Te.
  \end{cases}
\end{equation}
Similarly, substituting (\ref{press_dueto_force}) into (\ref{deft_to_press_mass}) and integrating with respect to $T$, yields the deformation of the elastic plate. Fig. \ref{figure3}a presents the fluid pressure-field at two instants during ($T=0.1, 0.9$) and after ($T=1.1, 2$) the application of external force of the form of (\ref{sudden_pe}), where $T_e=1$. During the application ($T<1$), the pressure in the impact locus decays with time and acts to resist the temporally constant external force. After the application period ($T>1$), the pressure at the locus instantaneously changes sign, now working to resist the plate's relaxation and increases with time. The insert in Fig. \ref{figure3}a focuses on pressure at the locus of application of the external force, given by
\begin{equation}\label{p_locus}
  P(\boldsymbol{X}=\boldsymbol{\bar{X}})=\frac{1}{T_e}\begin{cases}
    \frac{\Gamma(\frac{1}{3})}{12\pi}\frac{1}{T^\frac{1}{3}}, & T<Te\\
    \frac{\Gamma \left(\frac{1}{3}\right)}{12 \pi }\left(\frac{1}{T^\frac{1}{3}}-\frac{1}{(T-T_e)\frac{1}{3}}\right), & T>Te.
  \end{cases}
\end{equation} 
From (\ref{p_locus}), the rate of decay during application is $\sim T^{-1/3}$, and a discontinuity in pressure occurs at the instant of release (T=1) of the external force. 


Fig. \ref{figure3}b presents the deformation-field created in the elastic plate. While the deformation near the impact locus is negative, there is a region of significant positive deformation adjacent to the minima at the locus. An infinite number of maxima points of the deformations are expected from (\ref{G_d_to_G}). However,  all other maxima points are negligible in magnitude compared with the deformation near the locus. The maxima of this primary positive deformation region propagates radially both during and after the application of the external pressure field. The insert in Fig. \ref{figure3}b focuses on deformation at the locus, given by
\begin{equation}\label{GD}
  D(\boldsymbol{X}=\boldsymbol{\bar{X}})=\frac{1}{T_e}\begin{cases}
    \frac{\Gamma\left(-\frac{1}{3}\right)}{6}T^\frac{1}{3}, & T<Te.\\
    \frac{\Gamma\left(-\frac{1}{3}\right)}{6}\left(T^\frac{1}{3}-\left(T-T_e\right)^\frac{1}{3}\right), & T>Te.
  \end{cases}
\end{equation}
From (\ref{GD}), the deformation during force application is shown to increase as a power of $T^{1/3}$, and have a discontinuity in the rate of deformation at the moment of release of the external pressure ($T=1$). 



\begin{figure}
\centering
\includegraphics[width=0.7\textwidth]{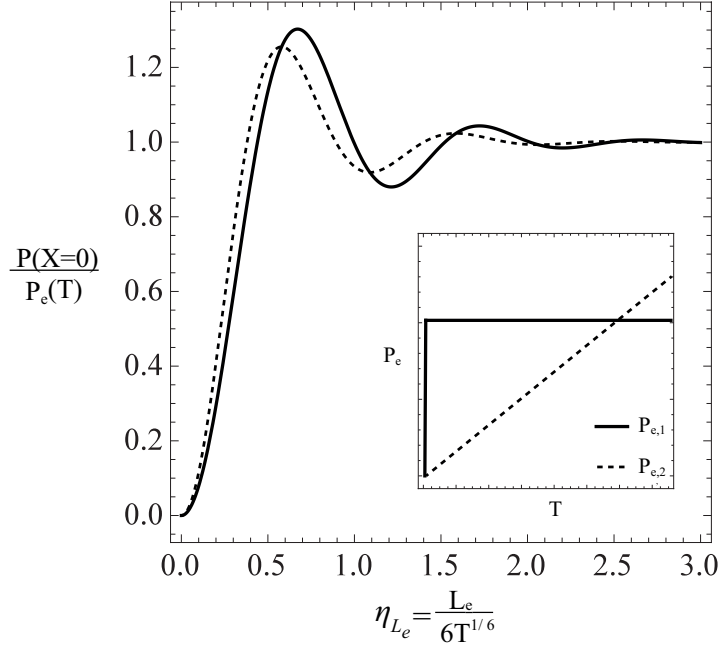}
\caption{The liquid pressure at $\boldsymbol{X}=0$ divided by the external pressure during application period vs. $\eta_{L_e}$. The solid line denotes the ratio of pressures as a result of an external pressure rapidly applied at $T=0$ and constant throughout the application period. The dashed line denotes the ratio of pressures as a result of an external pressure linearly increasing with time. Both external pressures are distributed evenly on a disk of radius $L_e$. The insert shows a schematic illustration of the evolution of the external pressure in time for both cases.}
\label{figure4}
\end{figure}

We now turn to explore the relation between the externally applied pressure field and the fluidic pressure field in order to examine impact mitigation properties of such configurations. This necessitates examination of finite external pressures, distributed both in spatially and temporally. We initially focus on a suddenly applied external pressure, uniform in both space and time with a time period $T_{e,1}$, spatial radius $L_e$ and a constant total impulse of $1$, given by
\begin{equation}\label{PE1}
P_{e,1}=\frac{1}{\pi L_e^2 T_{e,1}}\theta(L_e-|\boldsymbol{X}|)[\theta(T)-\theta(T-T_{e,1})].
\end{equation}
Convolving (\ref{G_p_to_G}) with (\ref{PE1}), the pressure ratio between the externally applied pressure and the fluidic pressure can be estimated for $T\leq T_{e,1}$ as
\begin{multline}\label{p_to_pe_rapid}
\frac{P(\boldsymbol{X}=0)}{P_{e,1}}=1-\, _0F_4\left(;\frac{1}{3},\frac{1}{3},\frac{2}{3},\frac{2}{3};
-\eta_{L_e}^6\right)\\-\frac{27}{2}  \eta_{L_e}^4 \Gamma \left(\frac{2}{3}\right) \, _0F_4\left(;1,\frac{4}{3},\frac{4}{3},\frac{5}{3};-\eta_{L_e}^6\right)+3 \eta_{L_e}^2 \Gamma \left(\frac{1}{3}\right) \, _0F_4\left(;\frac{2}{3},\frac{2}{3},1,\frac{4}{3};-\eta_{L_e}^6\right),
\end{multline}
where  $\eta_{L_e}=L_e/6T^{1/6}$. Eq. (\ref{p_to_pe_rapid}) is presented in Fig. \ref{figure4} (solid line). Three distinct periods are evident: I. An initial period of the impact, $2.5\lesssim\eta_{L_e}<\infty$, where the fluidic pressure closely follows the external pressure, II. The interval, $0.5\lesssim\eta_{L_e}\lesssim 2.5$, shows small oscillations of the pressure ratio going from mitigation to amplification and vise versa, and III. The period $0\lesssim\eta_{L_e}\lesssim 0.5$ where mitigation occurs and grows with time. However, since at the application of the suddenly applied external pressure, $ T=0^+$ and thus, $\eta_{L_e}\rightarrow\infty$ where $P(\boldsymbol{X}=0)\rightarrow P_{e,1}$, no mitigation can be achieved from external pressures where the rise time is order of magnitude smaller than the viscous elastic time scale.

\begin{figure}
\centering
\includegraphics[width=0.7\textwidth]{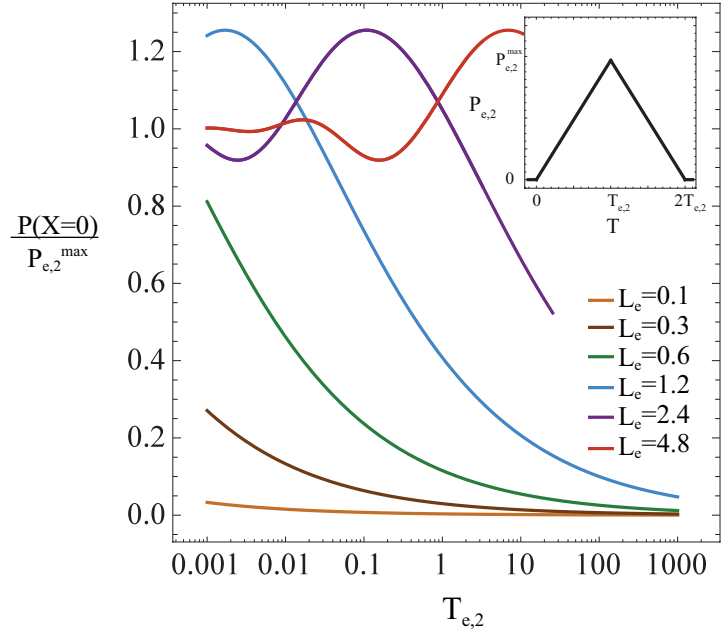}
\caption{Ratio of the liquid pressure at the center of impact to the external pressure, $P_{e,2}$, at the moment of maximal external pressure, $T=T_{e,2}$. External radii $L_e=0.1$, $0.3$, $0.6$, $1.2$, $2.4$, $4.8$ correspond to colors of orange, brown, green, blue, purple and red, respectively. The insert shows schematically the evolution of the external pressure in time.}
\label{figure5}
\end{figure}

Next, we turn to examine time-varying external pressures with rise time of order of magnitude of the viscous-elastic time scale. We model an external pressure, evenly distributed on a disk of radius $L_e$, linearly increasing in magnitude with respect to time until $T_{e,2}$ and than decreases linearly until vanishing at $2T_{e,2}$. The total impulse is 1, and $P_{e,2}$ is thus given by
\begin{equation}\label{p_lin}
P_{e,2}=\frac{\theta(L_e-|\boldsymbol{X}|)}{\pi L_e^2 T_{e,2}^2}[T\theta(T)-2(T-T_{e,2})\theta(T-T_{e,2})+(T-2T_{e,2})\theta(T-2T_{e,2})].
\end{equation}
Convolving (\ref{G_p_to_G}) with (\ref{p_lin}), denoting $\eta_{L_e}=L_e/6T^{1/6}$ and dividing by $P_{e,2}$ yields the ratio of pressures at the center for $T\le T_{e,2}$
\begin{equation}\label{press_to_pe_linear}
\frac{P(\boldsymbol{X}=0,T)}{P_{e,2}(T)}=\eta_{L_e}^6 G_{2,7}^{4,1}\left(\eta_{L_e}^6|
\begin{array}{c}
 0,1 \\
 -\frac{2}{3},-\frac{1}{3},0,0,-1,-\frac{2}{3},-\frac{1}{3} \\
\end{array}
\right)
\end{equation}
where $G_{2,7}^{4,1}$ is the Meijer G-function. The result is shown in Fig. \ref{figure4} with respect to similarity variable $\eta_{L_e}$ (dashed line) and with respect to the time where maximal external pressure is reached (i.e.  $T_{e,2}$) in Fig. \ref{figure5}. From Fig. \ref{figure5}, it is evident that for any width of external pressure $L_e$, mitigation may be achieved if the application time is sufficiently long. Specifically, for the case of $L_e=0.1$, mitigation of more than $90\%$ is achieved for external pressures applied over the period of $2T_{e,2}=10^{-3}$ or greater.

\appendix

\section{Results in dimensional form}
We present here some of the equations and results in dimensional form. These include the Green equation
\begin{equation}
\frac{\partial g}{\partial t}-\frac{h_0^3s}{12 \mu}\nabla^6 g=\delta(x)\delta(y)\delta(t),
\label{g_eq}
\end{equation}
the governing equation for deformation (\ref{D_gov})
\begin{equation}
\frac{\partial d}{\partial t}-\frac{h_0^3s}{12 \mu}\nabla^6 d=\frac{h_0^3}{12\mu}\nabla^2 p_e,
\label{gd_eq}
\end{equation}
the governing equation for pressure (\ref{P_gov})
\begin{equation}\label{gp_eq}
\frac{\partial p}{\partial t}-\frac{h_0^3s}{12 \mu}\nabla^6 p=\frac{\partial p_e}{\partial t},
\end{equation}
the Green function (\ref{g_eq}) 
\begin{equation}
g(\eta,t)=\left(\frac{12\mu}{h_0^3 s}\right)^\frac{1}{3}\frac{\Psi(\eta)}{(t-\bar{t})^{\frac{1}{3}}},
\end{equation}
the Green function for deformation (\ref{gd_eq})
\begin{equation}
g_d(\eta,t)=\left(\frac{h_0^3}{12\mu s^2 }\right)^\frac{1}{3}\frac{\Psi_d(\eta)}{(t-\bar{t})^{\frac{2}{3}}}
\end{equation}
(units are $[m/N\cdot s]$), the Green function for pressure (\ref{gp_eq})
\begin{equation}\label{gpa}
g_p(\eta,t)=\left(\frac{12\mu}{h_0^3 s}\right)^\frac{1}{3}\frac{\Psi_p(\eta)}{(t-\bar{t})^\frac{4}{3}}
\end{equation}
(units are $[Pa/N\cdot s])$, where $\eta$ for (\ref{gd_eq}-\ref{gpa}) is,
\begin{equation}
\eta=\left(\frac{12\mu}{h_0^3s}\right)^\frac{1}{6}\frac{|\boldsymbol{x}-\bar{\boldsymbol x}|}{6 (t-\bar{t})^\frac{1}{6}}.
\end{equation}
The deformation and pressure distribution as a result of a specific external pressure, $p_e=p_e(x,y,t)$, are obtained by the convolutions
\begin{equation}
d=\left(\frac{h_0^3}{12\mu}\right) g_d*p_e,\,\,\, p=g_p*p_e.
\end{equation}


\bibliographystyle{jfm}
\bibliography{Bib_File}

\end{document}